\documentclass{PoS}

\usepackage{lineno}

\title{Exploring the nature of 2HWC J2006+341 with HAWC and \textit{Fermi}-LAT}

\ShortTitle{Exploring the nature of 2HWC J2006+341}

\author{\speaker{Miguel Araya} for the HAWC Collaboration\footnote{for collaboration list see PoS(ICRC2019)1177}\\
        Universidad de Costa Rica \& \\Instituto Nacional de Astrof\'isica, \'Optica y Electr\'onica \\
        E-mail: \email{miguel.araya@ucr.ac.cr}}

\abstract{The High Altitude Water Cherenkov (HAWC) gamma-ray observatory is carrying out a detailed survey of the northern sky at TeV energies. 2HWC J2006+341 is a newly discovered source reported by the point source search in the 2HWC HAWC Observatory Gamma Ray Catalog. Using $\sim$1038 days of HAWC data we carried out a detailed analysis of the region revealing extended emission. No emission has been detected by other TeV instruments at the location of 2HWC J2006+341. We also analyzed publicly available data from the \textit{Fermi}-LAT that show for the first time the existence of an extended GeV source with a hard spectrum in the region of 2HWC J2006+341. Combined modeling of the data from HAWC and the \textit{Fermi}-LAT allowed us to explore different scenarios for the origin of 2HWC J2006+341.}

\FullConference{36th International Cosmic Ray Conference -ICRC2019-\\
		July 24th - August 1st, 2019\\
		Madison, WI, U.S.A.}

\begin{document}

\section{Introduction}
2HWC J2006+341 is a TeV gamma-ray source discovered by the High Altitude Water Cherenkov Observatory (HAWC) \cite{2017ApJ...843...40A} in the Cygnus Region and it is not associated to any other known TeV source. There are no known supernova remnants (SNRs) within one degree of the source's position \cite{2014BASI...42...47G} and the nearest pulsar in the ATNF catalog is PSR J2004+3429, located $0.4^{\circ}$ away. This pulsar is at a cataloged distance of 10.8 kpc. It has a spin-down power $\dot{E}=5.8\times10^{35}$ erg s$^{-1}$ and a characteristic age of 18 kyr \cite{2013MNRAS.435.2234B}. No pulsations have been found in \emph{Fermi}-LAT data around the location of PSR J2004+3429 \cite{2013MNRAS.435.2234B}.

The Fermi Large Area Telescope Fourth Source Catalog (4FGL) \cite{2019arXiv190210045T} shows two sources in the region. 4FGL J2004.3+3339, found $\sim0.64^{\circ}$ away from 2HWC J2006+341, is a point source associated to a binary system interacting with a molecular cloud (G70.7+1.2) \cite{kulkarni1992}. The other source, 4FGL J2005.8+3357, has a significance of $5.8\sigma$ in the 4FGL catalog, is located $0.23^{\circ}$ away from the reported position of 2HWC J2006+341 and has no association to any known source. No other gamma-ray sources are known in the region.

Follow-up observations by MAGIC and the \textit{Fermi}-LAT were only able to place upper limits in the direction of 2HWC J2006+341 \cite{2019MNRAS.485..356A}. The authors of this study noted that the offset from the HAWC emission and PSR J2004+3429 would require a pulsar kick velocity that is too high (>2300 km/s), which makes this pulsar unlikely to be related to the TeV emission.

In this contribution we present an updated analysis of HAWC and \textit{Fermi} data which reveals an extended region of gamma-ray emission at the location of 2HWC J2006+341 and discuss several scenarios for the origin of this emission.

\section{LAT data}
The \textit{Fermi} Large Area Telescope (LAT) is a converter/tracker telescope with sensitivity in the energy range between 20 MeV and $> 500$ MeV \cite{2009ApJ...697.1071A}. Pass 8 data were analyzed from the beginning of the mission (August 2008) to June 2019 with the latest response using the publicly available software \textit{fermitools}, version 1.0.1. SOURCE class events (including front and back interactions) were selected and standard cuts were applied. A spatial binning scale of 0.1$^{\circ}$ per pixel was used and the exposure was calculated using ten logarithmically spaced bins per decade in energy. Events within 20$^{\circ}$ around the coordinates (J2000) RA=$301.5^{\circ}$, Dec=$34^{\circ}$ were included in a maximum likelihood analysis with a base model consisting of the 4FGL catalog sources and diffuse emission and background templates (given by the files gll\_iem\_v06.fits and iso\_P8R2\_SOURCE\_V6\_v06.txt\footnote{See https://fermi.gsfc.nasa.gov/ssc/data/access/lat/BackgroundModels.html}). In order to carry out a more detailed analysis of the region near 2HWC J2006+341, the dim source 4FGL J2005.8+3357 was removed from the model since it is not associated to any known object. The normalizations of the 4FGL sources located less than $10^{\circ}$ around the position of 2HWC J2006+341 were left free in all the fits. The energy range chosen for the analysis was 5--500 GeV to lower the impact of the Galactic diffuse emission. The study of the region below 5 GeV is left for future work.

Excess emission was seen in a residual count map obtained after subtracting the best-fit model from the data (see Fig. \ref{fermi} below). In order to evaluate the significance of this emission, different spatial templates representing extended emission were compared to the single point source hypothesis. The templates used were the uniform disk and the 2D-gaussian as implemented in the LAT xml model format. The position of the center as well as the extension parameter (the radius for the disk and the sigma for the gaussian) were varied in steps of $0.1^{\circ}$ to maximize the likelihood in the grid. The test statistic (TS), defined as $-2\times$log$(L_0/L)$, with $L$ and $L_0$ the maximum likelihood values for a model with the additional source and without it (the null hypothesis), respectively, was calculated in each step. A simple power-law spectral assumption was used in all cases.

\section{HAWC data}
The HAWC Gamma-Ray Observatory is made of 300 optically isolated tanks of 7.3 m in diameter and 4.5 m in height, each containing 200,000 liters of water and equipped with 4 PMTs capable of detecting Cherenkov radiation from secondary particles. It is located in Sierra Negra, Mexico, at 4100 m a.s.l. \cite{2017ApJ...843...39A}. With a duty cycle > 95\%, an instantaneous field of view of 2 sr and a sensitivity to gamma-rays in the very high energy regime (> 300 GeV), it started operation in its full configuration in March 2015. The observations presented in this contribution contain a livetime of 1038.7 days. The data were binned according to the fraction of detectors hit and the estimated energy of the event using the ground parameter as defined in \cite{2019arXiv190512518H} applying standard gamma/hadron separation cuts and reconstruction \cite{2019arXiv190512518H}. Although the exposure time is $\sim2$ times that of the data presented in the 2HWC Catalog, note that there were more stringent cuts applied to the data set presented here, which were found to improve both the energy and angular resolutions \cite{2019arXiv190512518H}. The maximum likelihood technique was also used with HAWC data to find the spectral and morphological parameters of the sources in the model. A region of interest (RoI) with a radius of 8$^{\circ}$ around the location of 2HWC J2006+341 was fitted with a model containing the bright pulsar wind nebula (PWN) 2HWC J2019+367, a constant background extending over the RoI to account for unresolved sources and any possible contribution from diffuse emission, and 2HWC J2006+341.

\section{Results}
Using the LAT data, the maximum TS values obtained were 47.2 and 43.6 for the disk and gaussian templates, respectively, and a TS value of 24.0 was obtained for a point source at the location of 4FGL J2005.8+3357. The quantity TS$_{\mbox{\tiny template}}-$TS$_{\mbox{\tiny point}}$=$2\times$(log($L_{\mbox{\tiny template}}$)-log($L_{\mbox{\tiny point}}$)), related to the significance of extension, is greater than 16 in both cases, the typical threshold used to claim that a source is extended in validated LAT analyses \cite{2016ApJS..224....8A}. The best-fit radius of the disk was $0.82 \pm 0.05_{\mbox{\tiny stat}}^{\circ}$ and its center is located at the coordinates RA=301.50$^{\circ}$, Dec=34.35$^{\circ}$, with an estimated $1\sigma$ statistical uncertainty in the position of $0.1^{\circ}$. For the gaussian template, the resulting values for the radius ($1\sigma$) and the center coordinates were $0.6 \pm 0.1_{\mbox{\tiny stat}}^{\circ}$, RA=301.40$^{\circ}$, Dec=34.25$^{\circ}$, respectively ($1\sigma$ uncertainty of $0.15^{\circ}$). Both the disk and the gaussian templates provided similar improvements in the morphological description of the LAT data and so the gaussian template was chosen for the rest of the analysis for better comparison with HAWC data (see below). The spectrum of the emission in the 5--500 GeV energy range can be described with a simple power law ($\frac{dN}{dE} \propto E^{-\Gamma}$) with an index $\Gamma = 1.75 \pm 0.14_{\mbox{\tiny stat}}$ and an integrated flux of $(4.5 \pm 0.9_{\mbox{\tiny stat}})\times 10^{-10}$ cm$^{-2}$ s$^{-1}$. There was no significant improvement in the fit when a log parabolic power law function (logPar) was used to model the spectrum, the resulting increase in the TS with respect to the simple power-law was $\sim 1.0$.

In the HAWC analysis, fitting the location and spectrum of a point source gives a TS of 23.8. When fitting the data with a gaussian template and a disk template, the highest value was obtained for the gaussian (49.9 versus 36.3). The gaussian template fit also produced improved residuals and it was chosen for the rest of the analysis. The best fit radius of the gaussian ($1\sigma$) found was $0.73 \pm 0.14_{\mbox{\tiny stat}}^{\circ}$ and the coordinates of the center were RA=$301.56 \pm 0.18_{\mbox{\tiny stat}}^{\circ}$, Dec=$34.36 \pm 0.16_{\mbox{\tiny stat}}^{\circ}$. Using this morphology, we saw that the HAWC spectrum showed evidence of curvature at the $4\sigma$ level. The fit using a logPar of the form $$\frac{dN}{dE} = f_0 \left( \frac{E}{E_0} \right)^{\alpha - \beta \, \ln (E/E_0)}$$ with $E_0 = 7$ TeV resulted in a TS value of 67.0. The best-fit parameters found were $f_0 = (7.8_{-1.6}^{+2.1})\times 10^{-14}$ TeV$^{-1}$ cm$^{-2}$ s$^{-1}$, $\alpha = -3.1^{+0.4}_{-0.5}$ and $\beta = 1.0^{+0.8}_{-0.5}$ (the errors are all statistical). The energy range over which the source is detected is 2.5--14 TeV.

\begin{figure}
\begin{center}
\includegraphics[width=15.0cm,height=13.2cm]{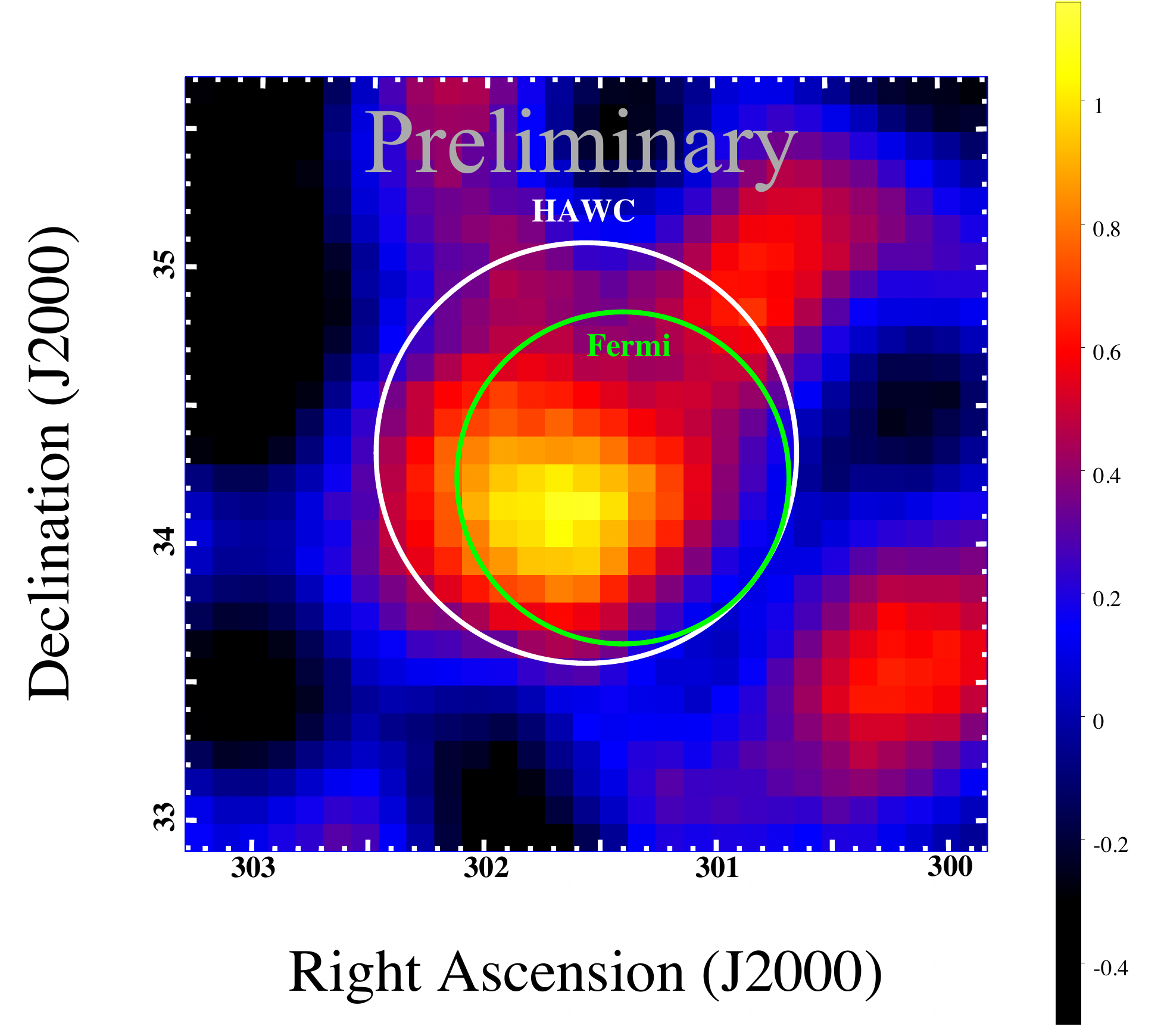}
\caption{Residual counts map obtained from LAT data (above 5 GeV) showing emission in the region of 2HWC J2006+341. The circles represent the extension of the best-fit gaussian templates found with HAWC (white) and \emph{Fermi} (green) data. The image was smoothed with a gaussian kernel with $\sigma = 0.25^{\circ}$.\label{fermi}}
\end{center}
\end{figure}

Fig. \ref{fermi} shows a LAT skymap with the residuals in the 5--500 GeV range obtained after subtracting the background model to the data. The map shows the extension of the best-fit gaussian templates obtained with the analysis of data from HAWC and the LAT.

\begin{figure}
\begin{center}
\includegraphics[width=12cm,height=9.3cm]{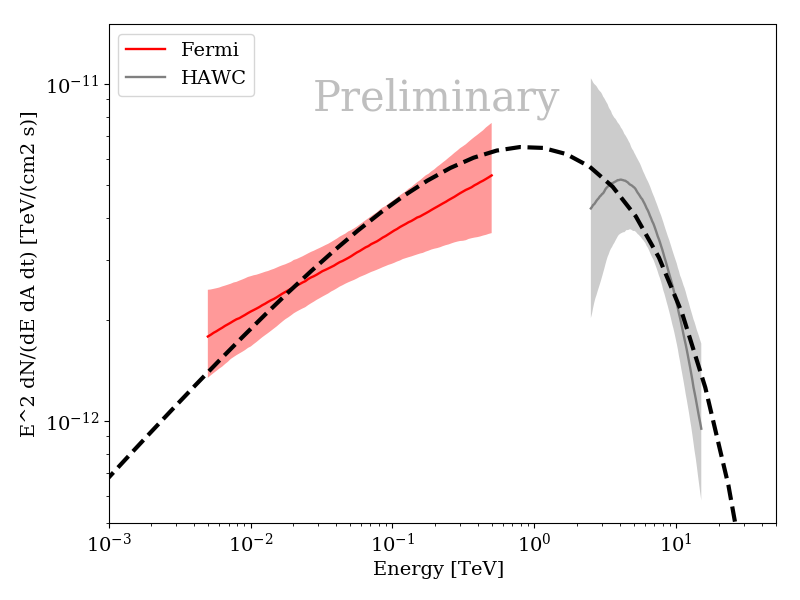}
\caption{SED of 2HWC J2006+341 showing the best-fit spectra and statistical error bands obtained with \emph{Fermi} (red) and HAWC (gray) data. The dashed line shows an IC/CMB emission model from high-energy electrons as explained in the text. \label{sed}}
\end{center}
\end{figure}

It is interesting to note that the extensions and locations of the sources that we found with both the HAWC and the LAT data are consistent within statistical uncertainties. The spectra measured by both instruments are also consistent. Fig. \ref{sed} shows the spectral energy distribution (SED) determined with the best-fit morphological templates found independently. Having a hard spectrum in the LAT regime and a soft spectrum in the TeV regime, it is clear that a cut off or a break should exist in between.

\section{Model}
Fig. \ref{sed} also shows a basic one-zone leptonic emission model resulting from inverse Compton (IC) scattering of CMB photons by high-energy electrons. The model was able to account for the overall spectrum. The particle energy ($\epsilon$) distribution used was a power-law with an exponential cutoff of the form $\epsilon^{-p} \,$e$^{-\epsilon/\epsilon_c}$ with $p=2.05$ and $\epsilon_c = 19$ TeV. The required total energy in the particles above 1 GeV, in terms of the source distance $d$, was $2.5 \times 10^{49} \left( \frac{d}{ \mbox{\small 10.8 kpc} } \right)^2$ erg.

A similar ``SED bump'' can be achieved in a hadronic scenario with the same distribution shape for a spectral index and a particle cutoff energy of 1.6 and 40 TeV, respectively. The total energy in the protons was $2.1 \times 10^{49} \left( \frac{d}{1 \, \mbox{\small kpc} } \right)^2  \left( \frac{1 \, \mbox{\small cm}^{-3}}{ n } \right)$ erg, in terms of the average density of the target material, $n$.

\section{Discussion}
The distance to the pulsar PSR J2004+3429 has been estimated to be $\sim11$ kpc. A rough calculation of the available rotational energy over the pulsar characteristic age is $\dot{E}\cdot 18$ kyr $\sim 3\times 10^{47}$ erg, which is much smaller than the energy required by the data shown in Fig. \ref{sed} (of the order of $10^{49}$ erg for a distance of 11 kpc). Therefore, unless the pulsar is actually much closer, it is not possible for PSR J2004+3429 to produce the GeV--TeV emission. Based on the angular size of the emission found here, the physical size of the emission region at a distance of 11 kpc would be $\sim 300$ pc, which is not realistic, at least for a source age of a few tens of kyr \cite{2018A&A...612A...2H}.

For a pulsar spin-down power $\dot{E}=5.8\times10^{35}$ erg s$^{-1}$, results by \cite{2018A&A...612A...2H} based on a PWN population study predict the associated PWN luminosity in the energy range 1--10 TeV of $\sim 5\times10^{33}$ erg s$^{-1}$ (without taking into account their relatively large model dispersion). Given the statistical uncertainties of the HAWC measurement presented here, if 2HWC J2006+341 is indeed the PWN associated to PSR J2004+3429, this translates to a source distance in the range 1.2--2.5 kpc. Of course, it would also be possible for a different, undetected, pulsar to produce the gamma-ray emission in the region of 2HWC J2006+341.

Another possibility is that the gamma-ray emission comes from the shell of a previously unknown supernova remnant (SNR). Several TeV shell candidates have been detected which have no known counterpart (or show very dim radio emission) at lower energies such as HESS J1912+101 and HESS J1614-518 \cite{2018A&A...612A...8H}, so it is possible that 2HWC J2006+341 is one more example of this subclass. The SNR scenario is supported by the similar angular sizes of the emission region seen by HAWC and \emph{Fermi}-LAT, which is generally not expected for a PWN due to the cooling of the particles as they propagate away from the pulsar. The energetics are also not a concern in the SNR scenario. Both leptonic and hadronic models require a total energy in the particles which is a small fraction of the typical kinetic energy in an SNR ($\sim 10^{51}$ erg) for a wide range of possible distances. In the hadronic scenario, the particle distribution required has a hard spectral index, which could result from cosmic rays interacting with dense target material distributed throughout the shell \cite{2014MNRAS.445L..70G}. The leptonic scenario also provides a natural explanation given the hard GeV spectrum.

In conclusion, 2HWC J2006+341 cannot be produced by PSR J2004+3429, unless the actual pulsar distance was much closer than determined by its dispersion measure ($\sim$1--2.5 kpc instead of 11 kpc). It could also be produced by another unknown pulsar. The PWN scenario might be in conflict with the similar GeV and TeV sizes found here. An SNR scenario for the origin of 2HWC J2006+341 is consistent with the data, although no lower energy counterparts are known. However, several SNR shell candidates have only been seen at TeV energies. Further observations at other wavelengths are encouraged for 2HWC J2006+341.

\acknowledgments
We acknowledge the support from: the US National Science Foundation (NSF); the US Department of Energy Office of High-Energy Physics; the Laboratory Directed Research and Development (LDRD) program of Los Alamos National Laboratory; Consejo Nacional de Ciencia y Tecnolog\'ia (CONACyT), M\'exico (grants 271051, 232656, 260378, 179588, 254964, 258865, 243290, 132197)(Cátedras 873, 1563, 341), Laboratorio Nacional HAWC de rayos gamma; L'OREAL Fellowship for Women in Science 2014; Red HAWC, M\'exico; DGAPA-UNAM (grants AG100317, IN111315, IN111716-3, IA102715, IN111419, IA102019, IN112218), VIEP-BUAP; PIFI 2012, 2013, PROFOCIE 2014, 2015; the University of Wisconsin Alumni Research Foundation; the Institute of Geophysics, Planetary Physics, and Signatures at Los Alamos National Laboratory; Polish Science Centre grant DEC-2014/13/B/ST9/945, DEC-2017/27/B/ST9/02272; Coordinaci\'on de la Investigaci\'on Cient\'ifica de la Universidad Michoacana; Royal Society-Newton Advanced Fellowship 180385. 
Thanks to Scott Delay, Luciano D\'iaz and Eduardo Murrieta for technical support. M.A. acknowledges support from Escuela de F\'isica at Universidad de Costa Rica and the Marie Sk\l{}odowska-Curie grant agreement No 690575 of the European Union's Horizon 2020 research and innovation programme. Thanks to Scott Delay, Luciano D\'iaz and Eduardo Murrieta for technical support.


\end{document}